\documentclass[12pt]{article}
\pdfoutput=1

\usepackage[bulletsep]{collref}
\usepackage{amssymb,graphicx}
\usepackage[intlimits]{amsmath}
\usepackage{bbm}
\usepackage[small]{subfigure}
\usepackage{makecell}
\usepackage{array}
\usepackage{booktabs}

\usepackage{MnSymbol}

\usepackage{tikz}
\usetikzlibrary {arrows.meta}
\usetikzlibrary{decorations.pathmorphing}
\usetikzlibrary{calc}
\usepackage{tikz-feynman}
\tikzfeynmanset{compat=1.1.0}

\makeatletter \@addtoreset{equation}{section} \makeatother

\makeatletter
\let\old@startsection=\@startsection
\let\oldl@section=\l@section
\renewcommand{\@startsection}[6]{\old@startsection{#1}{#2}{#3}{#4}{#5}{#6\mathversion{bold}}}
\renewcommand{\l@section}[2]{\oldl@section{\mathversion{bold}#1}{#2}}
\makeatother

\makeatletter
\let\old@makecaption=\@makecaption
\def\@makecaption{\small\old@makecaption}
\makeatother

\begin{document}



\renewcommand{\thefootnote}{\fnsymbol{footnote}}
\setcounter{footnote}{0}

\begin{center}
{\Large\textbf{\mathversion{bold} Wilson loops on the Coulomb branch \\
 of N=4 super-Yang-Mills }
\par}

\vspace{0.8cm}

\textrm{Jarne Moens$^{1}$ and
Konstantin~Zarembo$^{2,3}$}
\vspace{4mm}

\textit{${}^1$Theoretische Natuurkunde, Vrije Universiteit Brussel, Pleinlaan 2, \\ B-1050 Brussels, Belgium}\\
\textit{${}^2$Nordita, KTH Royal Institute of Technology and Stockholm University,
Hannes Alfv\'{e}ns V\"{a}g 12, 114 19 Stockholm, Sweden}\\
\textit{${}^3$Niels Bohr Institute, Copenhagen University, Blegdamsvej 17, \\ 2100 Copenhagen, Denmark}\\
\vspace{0.2cm}
\texttt{Jarne.Moens@vub.be, zarembo@nordita.org}

\vspace{3mm}

\par\vspace{1cm}

\textbf{Abstract} \vspace{3mm}

\begin{minipage}{13cm}
We study Wilson loops on the Coulomb branch of $N=4$ super-Yang-Mills theory, by solving for minimal surface that connect the contour on the boundary with the D3-brane in the bulk. The circular loop undergoes the Gross-Ooguri transition as a function of the radius and angular separation, and we fully map its phase diagram. As a byproduct we find evidence that the expectation value of the straight line is tree-level exact.
\end{minipage}

\end{center}

\vspace{0.5cm}


\setcounter{page}{1}
\renewcommand{\thefootnote}{\arabic{footnote}}
\setcounter{footnote}{0}

\section{Introduction}

Holography relates Wilson loops to minimal surfaces  with a given boundary \cite{Maldacena:1998im,Drukker:1999zq}. We are going to study Wilson loops on the Coulomb branch of the $\mathcal{N}=4$ super-Yang-Mills (SYM) theory, the problem that has not been addressed before. The Coulomb branch is dual to a D3-brane  in the bulk of $AdS_5\times S^5$ which adds a new element to the analysis of Wilson loops, because the minimal surface can now end on the D3-brane. A similar setup arises in defect CFTs \cite{Tai:2006bt,Drukker:2008wr,Nagasaki:2011ue,deLeeuw:2016vgp,Aguilera-Damia:2016bqv,Preti:2017fhw,Bonansea:2019rxh,Bonansea:2020rkv,Bonansea:2020vqq,Kristjansen:2024map}, where it leads to a number of interesting phenomena.

Minimal surfaces may undergo bifurcations as their boundary changes shape, and so do holographic Wilson loops. In this context the bifurcation is known as the Gross-Ooguri phase transition \cite{Gross:1998gk} which leads to a  non-analytic dependence of the Wilson loop on the geometric parameters. The Gross-Ooguri transition occurs in many configurations 
of holographic  Wilson loops
\cite{Rey:1998bq,Brandhuber:1998bs,Zarembo:1999bu,Olesen:2000ji,Kim:2001td,Park:2001kz,Tseytlin:2002tr,Drukker:2005cu,Ahn:2006px,Tai:2006bt,Nian:2009mw,Okuyama:2009fe,Burrington:2010yb,Dekel:2013kwa,Armoni:2013qda,Munkler:2018cvu}, and also has a weak-coupling counterpart \cite{Zarembo:2001jp,Correa:2018lyl,Correa:2018pfn}.

In the presence of a D-brane the Gross-Ooguri transition acquires a new twist as now the bifurcation occurs between a surface ending on the D-brane and a surface closing on itself. It was shown, for example, that
Wilson loops in the D3-D5 defect CFT feature an intricate phase structure  engendered by the D5-brane, both for rectangular \cite{Preti:2017fhw} and circular contours \cite{Bonansea:2019rxh}.
The presence of the D3-brane on  the Coulomb branch suggests the Gross-Ooguri transition may occur in this setup as well. This is the main question we are going to address in  this paper. 

Wilson loops on the Coulomb branch have not been studied before, and in  particular it is not known if they obey any non-renormalization theorems. 
The Coulomb branch preserves all the rigid supersymmetries, and those are usually responsible for the absence of quantum corrections.  One example is the trivial expectation value of the straight Wilson line in the conformal phase  \cite{Guralnik:2003di,Guralnik:2004yc}. The Wilson line may remain protected on the Coulomb branch as well and even though we will not attempt at a rigorous proof our strong-coupling calculations are consistent with the absence of radiative corrections in this case.

In sec.~\ref{sec:straight} we study  the Wilson line  and in sec.~\ref{sec:circular}  the circle. They are described by  Euclidean-signature minimal surfaces which have never been studied before in this context. On the contrary, there are well-known Minkowski solutions for spinning strings  on  the Coulomb branch \cite{Kruczenski:2003be,Espindola:2016afe}, which are quite important for the spectral problem \cite{Caron-Huot:2014gia}, amplitudes \cite{Alday:2025pmg} and one-point functions \cite{Skenderis:2006uy,Skenderis:2006di,Ivanovskiy:2024vel,Coronado:2025xwk}.

\section{Straight line}\label{sec:straight}

 A locally supersymmetric Wilson loop in the $\mathcal{N}=4$ theory is defined as a path-ordered exponential, with a coupling to scalars:
\begin{equation}
    W(\mathcal{C})= \mathop{\mathrm{tr}} \mathop{\mathrm{P}}\exp{\int_{\mathcal{C}}ds \left(iA_\mu \dot{x}^\mu + 
    |\dot{x}| n^i\Phi_i \right)}.
    \end{equation}
The scalar coupling is a unit six-dimensional vector $n ^i$ which may vary along the contour but here is taken  constant for simplicity. The AdS/CFT image of the Wilson loop is  a string anchored at the contour $\mathcal{C}$ on the boundary of $AdS_5$ and pinned to  the point  $n ^i$ on $S^5$ \cite{Maldacena:1998im}.

The scalar potential in the super-Yang-Mills theory  has flat directions along  diagonal matrices. Those are not lifted by quantum effects and allow scalars to take on expectation values partially breaking the $U(N+1)$ gauge symmetry. We consider the simplest setup with only one non-zero eigenvalue:
\begin{equation}\label{Higgs-cond}
\langle \Phi_i\rangle = 
    \begin{pmatrix}
       v_i & 0 & \cdots & 0 \\
0 & 0 & \cdots & 0 \\
\vdots & \ddots & \ddots & 0 \\
0 & 0 & 0 & 0
    \end{pmatrix}.
\end{equation}
The $U(N+1)$ gauge symmetry is then broken to $ U(N)\times U(1)$. The symmetry breaking is actually a small effect in the large-$N$ limit, as most of the gauge symmetry remains unbroken and the leading planar contribution to correlation functions still comes from the conformal $U(N)$ sector. 

The string dual of this setup is a single D3-brane floating in $AdS_5$ parallel to the boundary at the radial distance
\begin{equation}\label{z0-matching}
 z_0=\frac{\sqrt{\lambda }}{2\pi v}\,.
\end{equation}
The inverse proportionality between $z_0$ and $v$ follows from the dimensional analysis, while the precise coefficient comes from matching the energy of the string stretched to the horizon and the mass of a static W-boson. The R-symmetry orientation of the condensate $v_i/v$ determines where the D3-brane is placed on $S^5$.

\begin{figure}
\begin{center}
    
\begin{tikzpicture}[scale=0.70]
    \draw[->] (0,0) -- (5,0);
    \draw[->] (0,0) -- (0,5);
    \draw[->] (0,0) -- (2,2);

    \coordinate (lowA) at (2.5,0);
    \coordinate (lowB) at (4.5,2);
    \coordinate (upA)  at (2.5,3);
    \coordinate (upB)  at (4.5,5);
    \draw[yellow, thick] (lowA) -- (lowB);
    \draw[yellow, thick] (upA)  -- (upB);

    \fill[gray, opacity=0.4] (0,0) -- (2,2) -- (7,2) -- (5,0) -- cycle;
    \draw[gray!70!black, thick] (0,0) -- (2,2) -- (7,2) -- (5,0) -- cycle;

    \fill[teal, opacity=0.4] (0,3) -- (2,5) -- (7,5) -- (5,3) -- cycle;
    \draw[teal!70!black, thick] (0,3) -- (2,5) -- (7,5) -- (5,3) -- cycle;

    \def\stringcoords#1#2{
        (#1)
        ($(#1)!0.25!(#2) + (0.15,-0.15)$)
        ($(#1)!0.5!(#2)  + (-0.15,0.15)$)
        ($(#1)!0.75!(#2) + (0.15,-0.15)$)
        (#2)
    }

    \path let \p1 = (lowA), \p2 = (upA) in
        coordinate (start1) at (\x1,\y1)
        coordinate (end1)   at (\x2,\y2);
    \path let \p1 = (lowB), \p2 = (upB) in
        coordinate (start2) at (\x1,\y1)
        coordinate (end2)   at (\x2,\y2);

    \fill[yellow, opacity=0.1]
        plot [smooth] coordinates {\stringcoords{start1}{end1}}
        --
        plot [smooth] coordinates {\stringcoords{end2}{start2}}
        -- cycle;

    \draw[thick] plot [smooth] coordinates {\stringcoords{start1}{end1}};
    \draw[thick] plot [smooth] coordinates {\stringcoords{start2}{end2}};
    \node at (start1) {\textbf{\large $\cdot$}};
    \node at (end1)   {\textbf{\large $\cdot$}};
    \node at (start2) {\textbf{\large $\cdot$}};
    \node at (end2)   {\textbf{\large $\cdot$}};
    \node at (0,5.2) {z};
    \node at (5.5,-0.8) {AdS-Boundary};
    \node at (7,5.5) {D3-Brane};
\end{tikzpicture}
\caption{\label{stringWS}This figure depicts the string configuration of the Wilson loop where the contour is a straight line. The bottom gray plane represents the $AdS_5$-boundary, the green plane represents the D3-brane, the yellow shaded surface represents the worldsheet traced out by the string.}
\end{center}
\end{figure}
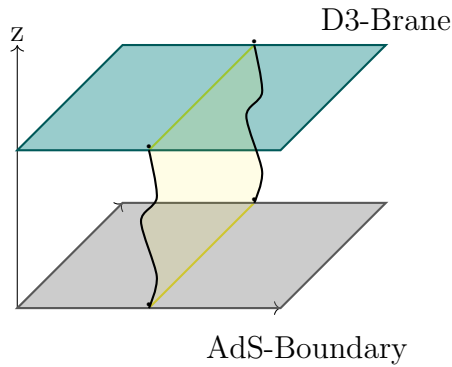

As we already mentioned,
the conformal sector overshadows the effects of symmetry breaking by the sheer volume of large-$N$ diagrams. To probe the effects of the symmetry breaking we introduce a connected correlator, in which the "trivial" conformal contribution is subtracted: 
\begin{equation}
 \left\langle W(\mathcal{C})\right\rangle_c\equiv \left\langle W(\mathcal{C})\right\rangle_{U(N+1)}
 -\left.\vphantom{\frac{1}{2}}\left\langle W(\mathcal{C})\right\rangle_{U(N)}\right|_{v=0}.
\end{equation}
In the string language, the subtraction term is the disc amplitude, proportional to  $N$ and decoupled from the D3-brane. The connected correlator  corresponds to the cylinder amplitude, of order $\mathcal{O}(1)$, with the string  attached by one end to the Wilson loop and by the other end to the brane, as illustrated in fig.~\ref{stringWS}.

The expectation value so defined depends on the 't~Hooft coupling $\lambda =g_{\rm YM}^2N$, the R-symmetry angle
\begin{equation}
 n^iv_i=v\cos\phi  ,
\end{equation}
and $vL$, the length of the contour measured in units of the Higgs condensate. In the ordinary perturbation theory quantum fields are expanded around the classical expectation value (\ref{Higgs-cond}). To the leading order in the weak-coupling expansion (at tree level), the field  $\Phi _i$ is just replaced by  $\left\langle \Phi _i \right\rangle$. For the Wilson loop this results in a perimeter law with a coefficient that depends on the relative R-symmetry orientation with respect to the Higgs condensate:
\begin{equation}\label{simple-straight}
 \left\langle W(\mathcal{C})\right\rangle_c= \,{\rm e}\,^{vL\cos\phi  }.
\end{equation}

The tree-level approximation receives loop corrections in the 't~Hooft coupling $\lambda =g_{{\rm YM}}^2N$. For a generic contour the loop corrections are certainly non-trivial. The  straight line however preserves half of the rigid supersymmetry and many cancellations may occur. A detailed  analysis is beyond the scope of our work, but it is easy to see that the first loop correction cancels: in the background Feynman gauge the vector potential and all the scalars have the same  propagators. Moreover the scalar propagator $\left\langle \Phi _i\Phi _j\right\rangle$ is proportional to $\delta _{ij}$ \cite{Alday:2009zm,Ivanovskiy:2024vel}  despite explicit R-symmetry breaking by the Higgs condensate. The loop diagram thus contains a kinematic factor $|\dot{x}_1||\dot{x}_2|n^in^j\delta _{ij}-\dot{x}_1^\mu \dot{x}_2^\nu \delta _{\mu \nu }$ identically vanishing for the straight line, before integration, and so the first correction of order $\lambda $ is indeed zero. We are not going to show that cancellations persists at higher orders, but will rather check that the classical perimeter law is reproduced by the string calculation at strong coupling.

The holographic calculations are valid in the opposite regime of $\lambda \rightarrow \infty $.
To the leading order at strong coupling, the correlator obeys the minimal area law, where the area is measured with the $AdS_5\times S^5$ metric (we display only one angle from $S^5$, the rest of the $S^5$ coordinates will never show up):
\begin{equation}
 ds^2=\frac{dx_\mu ^2+dz^2}{z^2}+d\theta ^2.
\end{equation}
Then,
\begin{equation}\label{W-generic}
 \left\langle W(\mathcal{C})\right\rangle_c\stackrel{\lambda \rightarrow \infty }{\simeq }\,{\rm e}\,^{-\frac{\sqrt{\lambda }}{2\pi }\,A(\mathcal{C})},
\end{equation}
where $A(\mathcal{C})$ is the regularized area of the minimal surface attached to the contour $\mathcal{C}$ on the boundary  at $z=0$ and ending on the brane at $z=z_0$, where is satisfies the usual Dirichlet-Neumann boundary conditions.

String theory easily reproduces (\ref{simple-straight}) at $\phi  =0$. The minimal surface is then a vertical wall crossing the brane at the right angle. The segment above the brane is to be chopped off. However, precisely this segment is the string image of a static W-boson used in the matching condition (\ref{z0-matching}). The matching identifies the action of the W-boson,  equal to $vL$, with the action of the semi-infinite string:
\begin{equation}
 vL\equiv \frac{\sqrt{\lambda }}{2\pi }\,A_{\rm above}=\frac{\sqrt{\lambda }}{2\pi }\,L\int_{z_0}^{\infty }\frac{dz}{z^2}
 =\frac{\sqrt{\lambda }}{2\pi }\,\,\frac{L}{z_0}\,.
\end{equation}
At the same time, the whole surface extending from the boundary to the horizon corresponds to the expectation value of the Wilson line at the conformal point, trivial due to supersymmetry protection. Hence $A_{\rm below}+A_{\rm above}=0$. This immediately gives $S_{\rm below}=-vL$ reproducing (\ref{simple-straight}) for the connected correlator on the Coulomb branch.

The case of  $\phi \neq 0$ requires a more elaborate calculation. The minimal surface in $AdS_5$ is the same but now the string moves on $S^5$ to compensate for misalignment between the Wilson loop and the Higgs condensate.  In the conformal gauge we can always choose
\begin{equation}
 x^0=\tau ,\qquad \theta =j\sigma, 
\end{equation}
and $z=z(\sigma )$. Here $j$ is an integration constant, which we later fix by imposing the boundary conditions. A similar proportionality constant in $x^0$ can be eliminated by a coordinate transformation $\tau \rightarrow \omega \tau $, $\sigma \rightarrow \omega \sigma $ that preserves the conformal gauge. 

The induced metric for this configuration is  
\begin{equation}
 ds^2=\frac{d\tau ^2+\left(\acute{z}^2+j^2z^2\right)d\sigma ^2}{z^2}\,.
\end{equation}
Here ans below  the prime denotes the derivative in $\sigma $: $\acute{z}=dz/d\sigma $.
The conformal gauge condition $ds^2=\,{\rm e}\,^{\chi }(d\tau ^2+d\sigma ^2)$ requires
\begin{equation}\label{line-Virasoro}
 \acute{z}^2+j^2z^2=1.
\end{equation}
This is enough to solve for $z$, giving:
\begin{equation}
 jz=\sin j\sigma.
\end{equation}
The boundary condition that the string subtends the angle $\phi$ exactly when it ends on the brane, $z(\phi /j)=z_0$, fixes the integration constant $j$:
\begin{equation}
 jz_0=\sin\phi .
\end{equation}

The metric of AdS blows up at the boundary producing a linear divergence in the area. The string action thus needs to be regularized. Conventional holographic prescription consists in cutting out a layer $z<\varepsilon $, subtracting a divergent perimeter term, and sending $\varepsilon$ to zero:
\begin{equation}
 A=\frac{1}{2}\int_{z(\tau ,\sigma )>\varepsilon }^{}d\tau \,d\sigma \,
 \left[
 \frac{(\partial_\sigma x_\mu )^2+(\partial_\sigma z)^2}{z^2}+(\partial_\tau \theta )^2
 \right]
 -\frac{L}{\varepsilon }\,.
\end{equation}
This prescription always gives finite renormalized area and hence defines a well-behaved expectation value for the Wilson loop    \cite{Drukker:1999zq}.
For the minimal surface at hand, the renormalized area is 
\begin{align}
 A=\frac{L}{2}\int d\sigma \,\,\frac{1+\acute{z}^2+j^2z^2}{z^2}-\frac{L}{\varepsilon }\,,
\end{align}
which can be simplified using (\ref{line-Virasoro}):
\begin{equation}
 A=L\int_{\varepsilon }^{z_0}\frac{dz}{z^2\sqrt{1-j^2z^2}}-\frac{L}{\varepsilon }\,.
\end{equation}
Evaluating the integral we get:
\begin{equation}
 A=\frac{L\sqrt{1-j^2z_0^2}}{z_0}=\frac{L\cos\phi }{z_0}\,.
\end{equation}
Taking into account the matching condition (\ref{z0-matching}) we recover the tree-level weak-coupling result (\ref{simple-straight}), for an arbitrary angle, from the holographic area law (\ref{W-generic}) at strong coupling.

It should be clear from this calculation that the perimeter law (\ref{simple-straight}) is a good approximation for any sufficiently large contour, satisfying $L\gg z_0$. The minimal surface is then approximately a cylinder $(x^\mu (\tau ),z(\sigma ))$, with $z(\sigma )$ just calculated. For a sufficiently large Wilson loop the curvature of the contour has little effect on the minimal surface because the brane sits close to the boundary. The  geometry of the minimal surface locally looks as if the contour were a straight line. The argument does not apply to contours with $L\sim z_0$ and for such contours the simple perimeter law will no longer hold. The Wilson loop will then depend non-trivially on the ratio $L/z_0$.
We are going to consider the simplest example of this type, a circle of radius $R$.

\section{Circular Wilson loop}\label{sec:circular}

The following coordinate transformation greatly facilitates the analysis:
\begin{equation}\label{g-alpha}
 z=R\,{\rm e}\,^{\alpha  },\qquad r=Rg\,{\rm e}\,^{\alpha  },
\end{equation}
where $r$ is the radial coordinate in the plane of the circle. The line element in these coordinates is
\begin{equation}
 ds^2=dg^2+2gdgd\alpha  +(1+g^2)d\alpha  ^2+g^2d\varphi ^2+d\theta ^2.
\end{equation}
The advantage of this coordinate system is that the scale invariance of the $AdS$ metric becomes a simple shift symmetry in $\alpha $.

In the conformal gauge we can take
\begin{equation}\label{phi-theta}
 \varphi =\tau ,\qquad \theta =j\sigma ,
\end{equation}
with $g$ and $\alpha $ depending on $\sigma $ only. Here, again, $j$ is an integration constant, to be fixed later by imposing the boundary conditions.

The string action evaluated  on  this solution is
\begin{equation}\label{stringac}
 S=\frac{\sqrt{\lambda }}{2}\int_{}^{}
 d\sigma \,
 \left[
 \acute{g}^2+2g\acute{g}\acute{\alpha }+(1+g^2)\acute{\alpha  }^2+g^2+j^2
 \right]-\frac{\sqrt{\lambda }\,R}{\varepsilon }\,.
\end{equation}
The boundary of $AdS_5$ in the coordinates we are using is reached at $\alpha  \rightarrow -\infty $, $g\rightarrow \infty $ with $g\,{\rm e}\,^{\alpha  }\rightarrow 1$. The cutoff on the holographic coordinate $z>\varepsilon $ translates into the cutoff on $\alpha $ and $g$ such that
\begin{equation}
 g_{\rm max}=\frac{R}{\varepsilon }\,.
\end{equation}

The equations of motion constitute a system of two second-order ODEs on $g$ and $\alpha $. Instead of solving them directly we will find two first integrals, which is enough to reduce the system to quadratures. One  integral of motion follows from the conformal gauge condition which
 boils down to
\begin{equation}\label{Virasoro}
 \acute{g}^2+2g\acute{g}\acute{\alpha  }+(1+g^2)\acute{\alpha  }^2+j^2=g^2.
\end{equation}
Substituting this condition into (\ref{stringac})  considerably simplifies the string action:
\begin{equation}\label{simplified-string-action}
 S=\sqrt{\lambda }\int_{}^{}d\sigma \,g^2-\sqrt{\lambda }\,g_{\rm max}.
\end{equation}

To find the second integral of motion we further exploit the symmetries of the problem. The shift symmetry of the string action (\ref{stringac}) implies conservation of the dilaton charge
\begin{equation}\label{dilaton}
 \sqrt{\epsilon }=(1+g^2)\acute{\alpha  }+g\acute{g}.
\end{equation}
Yet another integration constant, the dilaton charge will  be later fixed by the boundary conditions. The reason for denoting it by $\sqrt{\epsilon }$ will become clear shortly, $\epsilon $ should  not be confused with the regulator parameter $\varepsilon $.

Excluding $\acute{\alpha }$ from the two integrals of motion yields a first-order differential equation for $g$ which can be integrated in elliptic functions. It is instructive to start with the simpler case of $j=0$ that corresponds to the Wilson loop aligned with the Higgs condensate on $S^5$. The string then moves only in $AdS_5$.

\subsection{Aligned configuration}\label{sec:aligned}

Solving (\ref{dilaton}) for $\acute{\alpha }$ and substituting the result in (\ref{Virasoro}) we find (setting also $j=0$):
\begin{equation}\label{energy-conservation}
 \acute{g}^2-g^2-g^4=-\epsilon .
\end{equation}
This equation lends itself to a lucid mechanical analogy. The left-hand side is the energy of a particle moving in an upside-down quartic potential. Starting at infinity, the particle climbes the potential until it runs out of steam and rolls back. Unless the energy is exactly zero, whence plodding uphill continues ad infinitum. Overshooting the hilltop is not possible because $\epsilon >0$ by default -- we defined the dilator charge as $\sqrt{\epsilon }$.

At zero energy ($\epsilon =0$) the equations are solved in elementary functions:
\begin{equation}\label{hemi-g}
    g=\frac{1}{\sinh\sigma}.
\end{equation}
The dilaton charge conservation (\ref{dilaton}) can be also integrated, determining the shape of the minimal surface:
\begin{equation}
    z=R\tanh\sigma,\quad r=\frac{R}{\cosh\sigma}\,.
\end{equation}
This is the well-known hemisphere ($z^2+r^2=R^2$) solution  \cite{Berenstein:1998ij,Drukker:1999zq}, a minimal embedding of 
$AdS_2$ into $AdS_5$. 

The surface closes on itself and thus describes the expectation value of the circular Wilson loop in the conformal phase. The string action evaluated by integrating (\ref{simplified-string-action}) gives
\begin{equation}
 S_{\rm hemisphere}=-\sqrt{\lambda }\,.
\end{equation}
This well-known result determines the expectation value of the circular Wilson loop in the conformal phase, at strong coupling: $W\simeq \,{\rm e}\,^{\sqrt{\lambda }}$. The predictions can be confronted with the direct resummation of Feynman diagrams \cite{Erickson:2000af,Drukker:2000rr}, justified by  localization of the path integral \cite{Pestun:2007rz}. The successfull comparison between of the diagram resummation and the string calculation provides the basic example of weak to strong coupling interpolation in AdS/CFT.

\begin{figure}[t]
\begin{center}
 \subfigure[]{
   \includegraphics[height=4.5cm] {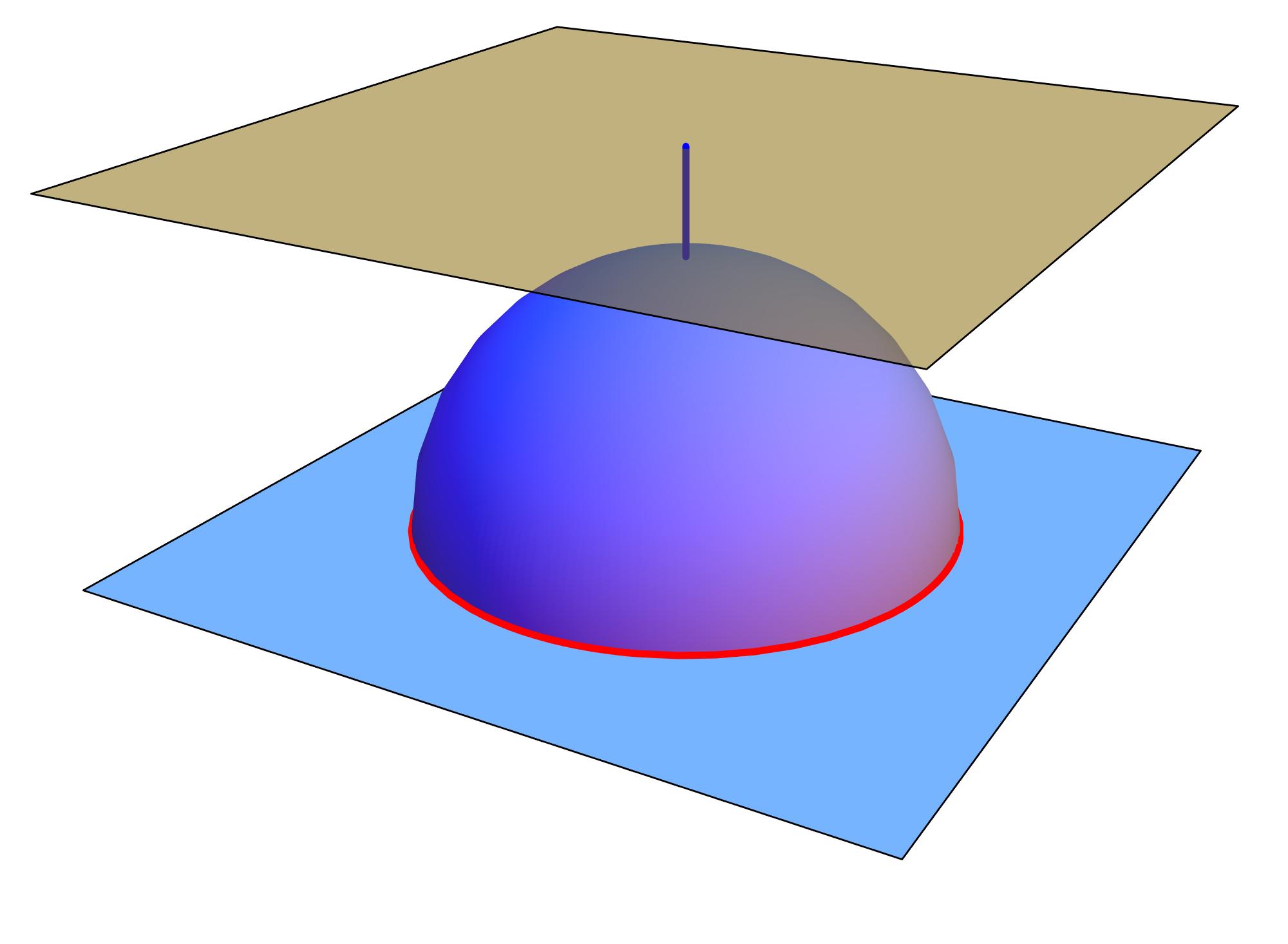}
   \label{fig2:subfig1}
 }
 \subfigure[]{
   \includegraphics[height=4.5cm] {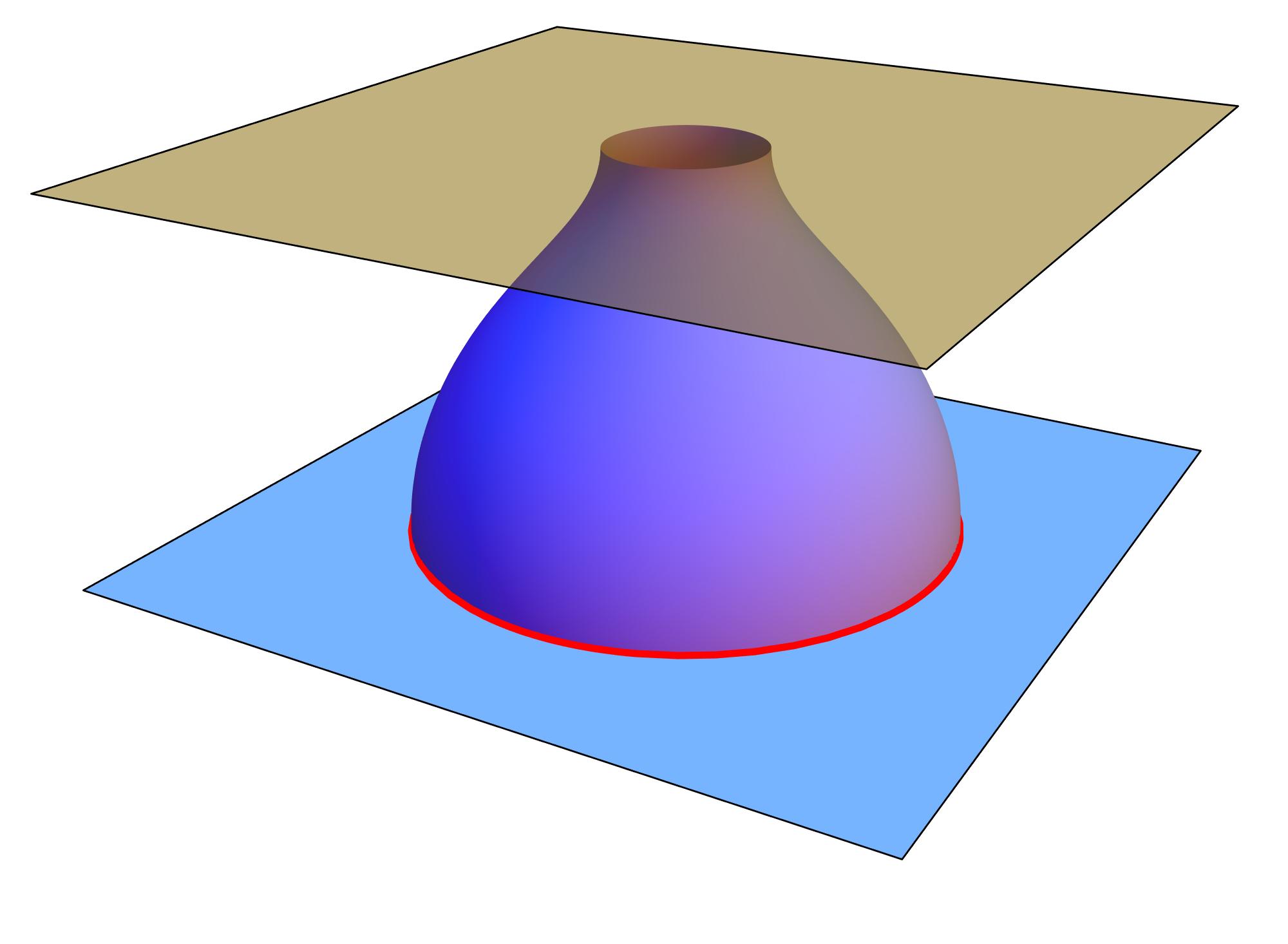}
   \label{fig2:subfig2}
 }
\caption{\label{minsurfaces}\small Two types of minimal surfaces conrtributing to the string path integral: (a) the hemisphere, connected to the brane by an infinitely thin tube, and (b) cylindrical minimal surface with a neck.}
\end{center}
\end{figure}

The hemisphere solution is also relevant for the Coulomb branch, even if it does not obey the right boundary conditions by itself. The string can be connected to the brane by an infinitely thin tube or, more precisely, by a supergravity propagator \cite{Berenstein:1998ij}, fig.~\ref{fig2:subfig1}. The hemisphere thus constitutes a legitimate saddle-point of the string path integral with the cylinder topology\footnote{Surfaces of wrong topology do contribute to the string path integral, by the mechanism just described. A clear manifestation can be found in a
 flat-space example of \cite{Gross:1998gk,Zarembo:1999bu}, where the exact quantum amplitude can be explicitly evaluated \cite{Zarembo:1999bu}.}. Another possibility is a minimal surface with the topology of a cylinder connected to the brane by a neck of finite width, fig.~\ref{fig2:subfig2}. The string path integral picks the dominant contribution from the saddle with the smaller area, and which one of the two solutions is the true minimum depends on the parameters of the problem. The competition between the two saddles is the essence of the Gross-Ooguri transition.

For the connected minimal surface the boundary conditions on the brane require
\begin{equation}
 \acute{r}=0~{\rm when}~z=z_0.
\end{equation}
Upon the change of variables (\ref{g-alpha}) and using (\ref{dilaton}) this condition becomes $\acute{g}=-\sqrt{\epsilon }\,g$. Taking into account the energy conservation (\ref{energy-conservation}), the ultimate value of $g$, where the evolution stops and the surface hits the brane, is 
\begin{equation}\label{max-g}
 g_*=\sqrt{\epsilon }\,.
\end{equation}

The equation of motion (\ref{energy-conservation}), and subsequently (\ref{dilaton}) can be solved in elliptic functions, giving rise to an explicit representation of the minimal surface in the conformal gauge. The explicit form of the solution and the derivation are presented in the appendix~\ref{explicit-elliptic:sec}, because we do not need these explicit results for evaluating the area. It is sufficient to use the equations of motion in the first-order form.

The idea is to change variables from $\sigma $ to $g$, the Jacobian of which is completely determined by (\ref{energy-conservation}). In the $g$-variables the string action (\ref{simplified-string-action})  becomes
\begin{equation}\label{integral-form-action}
 S=\sqrt{\lambda }\int_{g_*}^{g_{\rm max}}\frac{dg\,g^2}{\sqrt{g^2+g^4-\epsilon }}-\sqrt{\lambda }\,g_{\rm max}.
\end{equation}
Regularization can be removed by subtracting the linearly growing term from the integrand, resulting in a finite elliptic integral:
\begin{equation}\label{ell-action}
 S=-\sqrt{\lambda }\int_{\sqrt{\varepsilon }}^{\infty }dg\,
 \left(1-\frac{g^2}{\sqrt{g^2+g^4-\epsilon }}\right)-\sqrt{\lambda \epsilon }\,.
\end{equation}

The elliptic modulus of the integral is pure imaginary\footnote{Reflecting perhaps the Euclidean nature of the solution,  for spinning strings the modulus is typically real.}.
It is useful to introduce a new variable 
\begin{equation}\label{kappa-epsilon}
 \kappa^2 =\frac{\sqrt{1+4\epsilon }-1}{\sqrt{1+4\epsilon }+1}\,,
\end{equation}
 taking values in the interval $(0,1)$.
Equivalently,
\begin{equation}\label{epsilon-kappa}
 \epsilon =\frac{\kappa ^2}{(1-\kappa ^2)^2}\,.
\end{equation}
The modulus of the elliptic integral in (\ref{ell-action}) is given by\footnote{In the physics literature two different conventions are used for the elliptic integrals, one is adopted in {\tt Mathematica} in the functions like {\tt EllipticK[m]}, another is found in the standard reference, e.g. \cite{gradshteyn2014table}.
The two conventions are simply related, to avoid confusion we list both.}
\begin{equation}\label{ell-modulus}
 m=-\kappa ^2~({\rm \tt Mathematica})\qquad k=i\kappa ~({\rm \tt Gradshtein-Ryzhik}~\cite{gradshteyn2014table}).
\end{equation}
We will ancounter other elliptic functions, always with the same modulus which we therefore not indicate explicitly. The standard elliptic integrals of the 1st, 2nd and the 3rd kind will be denoted by $F(\varphi )$, $E(\varphi )$ and $\Pi (n,\varphi )$. 

The string action, in these notations, integrates to
\begin{equation}\label{string-action-elliptic}
 S=-i\sqrt{\frac{\lambda }{1-\kappa ^2}}
 \left(E(is_*)-F(is_*)\right)-\sqrt{\lambda\epsilon }\,,
\end{equation}
where 
\begin{equation}\label{sstar}
 \cosh  s_*=\frac{1}{\kappa }\,.
\end{equation}
The whole answer is of course real inspite of the imaginary $i$ appearing here and there.

The action as written is a function of an auxiliary parameter $\epsilon $, and we would like to express the action through the geometric data, in this case the radius of the circle and the position of the brane that only enter through their ratio $z_0/R$. To this end, we can use the boundary conditions on $z(\sigma )$  together with representation (\ref{g-alpha}). The conservation law (\ref{dilaton}) divided by $1+g^2$ and integrated from $g_*=\sqrt{\epsilon }$ to infinity gives $\alpha $ evaluated at the position of the brane. The boundary conditions require that this equals $\ln (z_0/R)$. Introducing a useful notation:
\begin{equation}\label{v-integral-form}
 v_*=\sqrt{\epsilon }\int_{\sqrt{\epsilon }}^{\infty }\frac{dg}{(1+g^2)\sqrt{g^2+g^4-\epsilon }}\,,
\end{equation}
the result can be compactly written as
\begin{equation}\label{z0}
 \frac{z_0}{R}=\frac{\,{\rm e}\,^{v_*}}{\sqrt{1+\epsilon }}
\end{equation}
In terms of the standard elliptic integrals,
\begin{equation}\label{vstar-elliptic}
 v_*=\frac{i\kappa }{\sqrt{1-\kappa ^2}}\left(
 \Pi (1-\kappa ^2,is_*)-F(is_*)
 \right).
\end{equation}
The last two equations, together with (\ref{string-action-elliptic}) and (\ref{sstar}), express the string action as a function of $R/z_0$ in a parameteric form.

\begin{figure}[t]
\begin{center}
 \subfigure[]{
   \includegraphics[width=6.5cm] {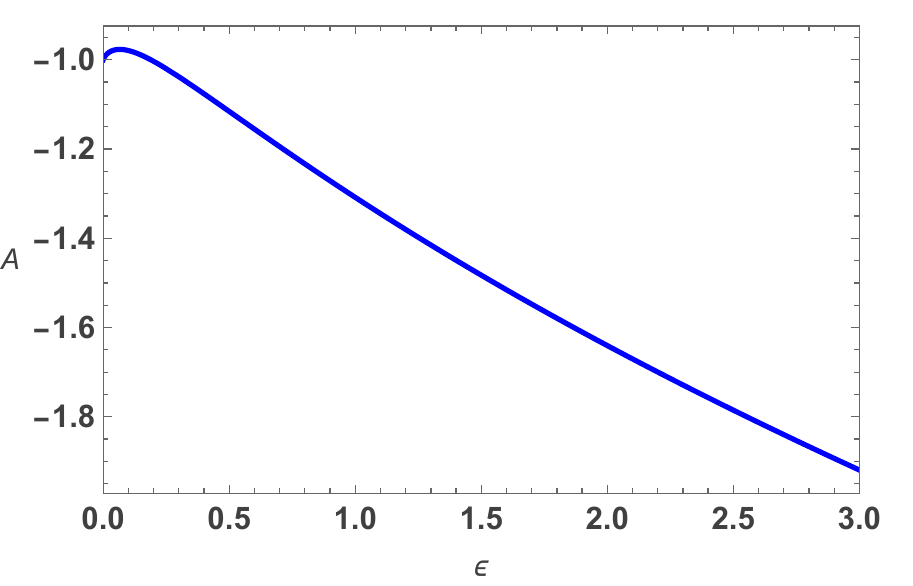}
   \label{fig3:subfig1}
 }
 \subfigure[]{
   \includegraphics[width=6.5cm] {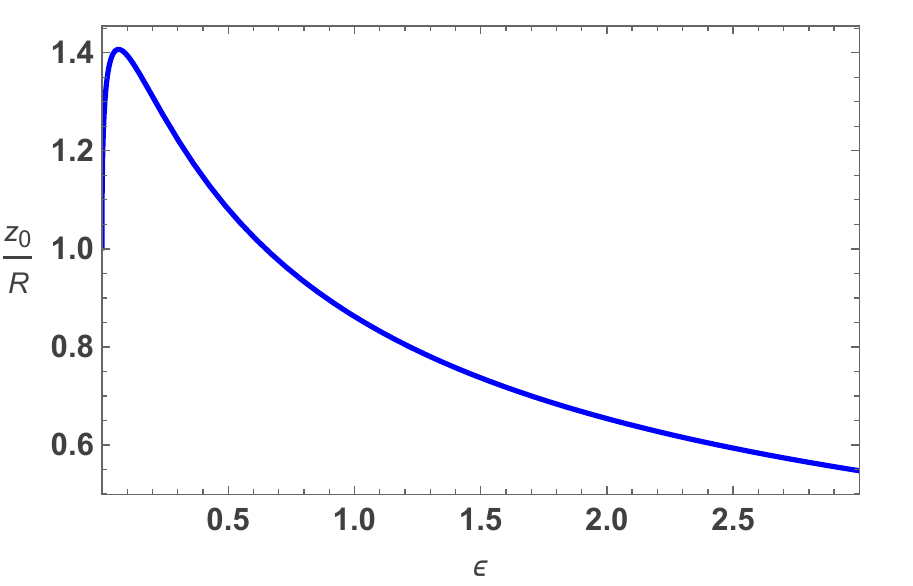}
   \label{fig3:subfig2}
 }
\caption{\label{area-and-z0/R}\small  The relation between the physical parameters of the string configuration and the energy $\epsilon $: (a) 
the worldsheet area; (b) the ratio of the bulk position of the brane $z_0$ and the radius $R$ of the circle. The results for the connected surface
smoothly match to the hemisphere solution zero energy. 
}
\end{center}
\end{figure}

The worldsheet area (the action divided by $\sqrt{\lambda }$) as a function of the auxiliary variable $\epsilon $ and the ratio $z_0/R$ are plotted in fig.~\ref{area-and-z0/R}. Both are non-monotonic functions of $\epsilon$ that initially grow and then decrease. It is easy  to understand why. Small energy corresponds to a hemisphere with an infinitesimal neck attached to it, the neck obviously  adds to the area and extends the surface's reach into the bulk. The area is then bigger than that of the hemisphere and the hight (the value of $z_0$) increases with $\epsilon $ growing. It is also clear why the connected surface cannot reach arbitrary high latitudes. The spacial extent of the Wilson loop counteracts the tendency of the neck to shrink only to a certain degree and for a fixed $R$ there is an upper bound on $z_0$ that the surface can reach before collapsing. The hole in the surface becomes wider and wider  with  lowering $z_0$, as should be clear from fig.~\ref{fig2:subfig2}, eating up area and eventually the surface can dimish its area by lowering $z_0$.

\renewcommand{\arraystretch}{0.98}  
\setlength{\tabcolsep}{3pt}         
\begin{table}[t]
  \centering
  \makebox[\textwidth][c]{   
  \begin{tabular}{|
      c|
      >{\centering\arraybackslash}p{5.5cm}|
      >{\centering\arraybackslash}p{5.5cm}|
    }
    \hline
    Ratio 
      & Connected solution 
      & Hemisphere \\
    \hline

    $\frac{z_0}{R} > 1.456$ 
      & \makecell{Does not exist \\ ~ } 
      & \makecell{Dominant} \\
    \hline

    $1.456 > \frac{z_0}{R} > 1$ 
      & \makecell{Two branches \\ stable + unstable } 
      & \makecell{ } \\
      
    \hline

    $\frac{z_0}{R} < 1.322$ 
      & \makecell{ Dominant \\ ~} 
      & \makecell{Subdominant} \\

    \hline

    $ \frac{z_0}{R} < 1$ 
      & \makecell{One branch \\ ~} 
      & \makecell{} \\
      
    \hline

  \end{tabular}
  }
  \caption{\label{minarea-table}\small The structure of the minimal surface for various ranges of parameters.}
\end{table}

The maximum in area is reached at $\epsilon _{\rm max}=0.0658$ simultaneously with the maximum in hight, corresponding to $z_{\rm max}=1.407R$. Between $z_0=R$ and $z_0=z_{\rm max}$  two connected solutions exist, one stable another unstable, which   meet and annihilate at $z_0=z_{\rm max}$. The area of the unstable solution is obviously bigger but even the stable branch does not necessarily realize the global minimum of the action. The area of the stable solution crosses the hemisphere line at $\epsilon _{\rm c}=0.1882$, equivalent to $z_{\rm c}=1.322R$. Between $z_{\rm c}$ and $z_{\rm max}$ the dominant saddle-point is the hemisphere with an infinitely thin tube attached, fig.~\ref{fig2:subfig1}. 
The structure of the minimal surface is summarized in the table~\ref{minarea-table}.

\begin{figure}[t]
 \centerline{\includegraphics[width=13cm]{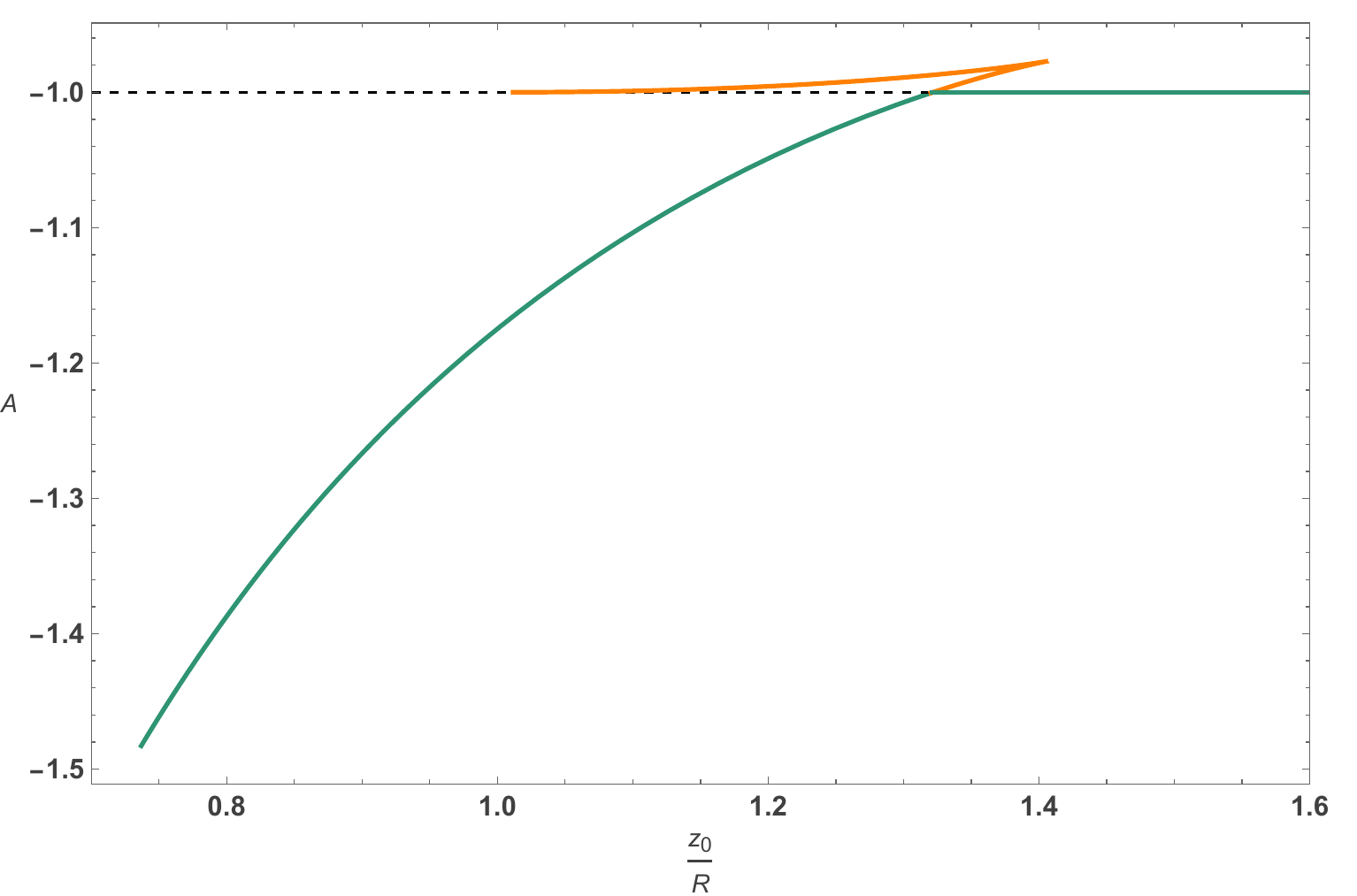}}
\caption{\label{minarea}\small The area of the minimal surface as computed from (\ref{string-action-elliptic}), (\ref{sstar}), (\ref{z0}) and (\ref{vstar-elliptic}). 
The two curves with a swallow-tail junction at $z_0=z_{\rm max}$ represent the minimal surface with the neck. It has two branches, stable (the lower curve) and unstable (the upper curve).
The  horisontal line at -1 is the area of the hemisphere.
The green line is the true minimum of the string action that switches between the hemishere and the surface with a neck at $z_0=z_{\rm c}$.}
\end{figure}

In fig.~\ref{minarea} we plot the area of the minimal surface given by (\ref{string-action-elliptic}), (\ref{sstar}), (\ref{z0}), (\ref{vstar-elliptic}). 
We see that the competition between connected and disconnected solutions leads to the Gross-Ooguri transition at the critical radius
\begin{equation}
 R_{\rm c}=0.7566z_0.
\end{equation}
The transition is first-order, with discontinuous first derivative at the transition point, and the existence of the "overcooled" region in the wrong phase.

The Wilson loop of  large radius, or  small $z_0/R$, corresponds to
large energy $\epsilon$. The complicated integrals in (\ref{integral-form-action}), (\ref{v-integral-form}) are then subleading and can be dropped:
\begin{equation}
 S\stackrel{\epsilon \rightarrow \infty }{\simeq }-\sqrt{\lambda \epsilon }\,,
 \qquad 
 \frac{z_0}{R}\stackrel{\epsilon \rightarrow \infty }{\simeq }\frac{1}{\sqrt{\epsilon }}\,,
\end{equation}
which upon identification (\ref{z0-matching}) gives
\begin{equation}
 S\simeq -2\pi Rv.
\end{equation}
This recovers the simple perimeter law (\ref{simple-straight}), with $\phi =0$ in this case. As mentioned earlier a large Wilson loop with $Rv\gg 1$ behaves approximately as a straight line, with curvature corrections suppressed by $1/R$. The explicit  calculation for the circle  confirms this expectation.

The curvature correction can be readily calculated by pushing the expansion in $1/\epsilon $ to higher orders: 
\begin{equation}
 S=-\sqrt{\lambda \epsilon }\left(1+\frac{1}{3\epsilon }-\frac{1}{35\epsilon ^2}+\ldots \right),\qquad 
 \frac{z_0}{R}=\frac{1}{\sqrt{\epsilon }}\left(1-\frac{1}{6\epsilon }
 +\frac{89}{2520\epsilon ^2}+\ldots 
 \right)
\end{equation}
It is easy to see that the parameter of this expansion is effectively
\begin{equation}
 \frac{z_0^2}{R^2}=\frac{\lambda }{4\pi ^2R^2v^2}\,,
\end{equation}
making the result look as perturbative series:
\begin{equation}
 \ln \left\langle W\right\rangle_c=2\pi Rv
 \left(1+\frac{\lambda }{24\pi ^2R^2v^2}
 +\frac{17\lambda ^2}{40320\pi ^4R^4v^4}+\ldots \right).
\end{equation}

The emergence of perturbative series at strong coupling is a well-known phenomenon first reported in \cite{Berenstein:2002jq}. For Wilson loops it was found in various incarnations of defect CFT \cite{Bonansea:2020vqq,Kristjansen:2024map}, a setup in many respects
similar to ours,
and also for correlators of Wilson loops with local operators in the large-charge limit \cite{Zarembo:2002ph,Pestun:2002mr} where the string solutions carry certain visual similarities to the ones we are studying \cite{Pestun:2002mr}. In all these cases it was possible to identify the diagrams of perturbation theory that reproduce the strong-coupling result, thus enabling a direct comparison between planar diagrams and string theory. We expect a similar story to unfold here. The relevant diagrams are presumably rather simple, most probably tree-like, as the coefficients of the expansion are simple rational numbers. We leave this interesting question for future work.

\subsection{Misaligned configuration}

In the  previous section we studied the minimal surface with no extent in $S^5$ -- that corresponds to setting the integration constant $j$ in (\ref{phi-theta}) to zero. Here we consider a more general solution with a non-zero $j$, and consequently non-trivial motion on $S^5$. The extension to non-zero $j$ is straightforward, and we will skip most technical details of the calculation as it follows basically the same steps.  

Eliminating $\acute{\alpha }$ from the conservation laws  (\ref{Virasoro}) and (\ref{dilaton}), we find for the energy of the analog mechanical system:
\begin{equation}\label{new-energy-conservation}
 \acute{g}^2-(1-j^2)g^2-g^4+j^2=-\epsilon ,
\end{equation}
a generalization of 
(\ref{energy-conservation}). 
The boundary condition (\ref{max-g}) now becomes
\begin{equation}\label{new-g*}
 g_*=\sqrt{\epsilon +j^2}\,.
\end{equation}
The equations are integrated in elliptic functions, and we again relegate all explicit formulas the the appendix~\ref{explicit-elliptic:sec}.

The elliptic modulus is governed by the parameter
\begin{equation}
 \kappa^2 =\frac{\sqrt{(1+j^2)^2+4\epsilon }-1+j^2}{\sqrt{(1+j^2)^2+4\epsilon }+1-j^2}\,,
\end{equation}
which now takes values between $j^2$ and $1$. Its relation to the moduli of the elliptic function is the same (\ref{ell-modulus}) as before.
The inverse relation is
\begin{equation}
  \epsilon =\frac{(\kappa ^2-j^2)(1-\kappa ^2j^2)}{(1-\kappa ^2)^2}\,.
\end{equation}

The action, as before, is given by (\ref{simplified-string-action}). After changing the integration variable to $g$ and removing regularization the action becomes
\begin{equation}\label{string-action-integral-j}
 S=-\sqrt{\lambda }\int_{\sqrt{\epsilon +j^2}}^{\infty }dg\,
 \left[
 1-\frac{g^2}{\sqrt{(1-j^2)g^2+g^4-j^2-\epsilon }}
 \right]
 -\sqrt{\lambda (\epsilon +j^2)}\,,
\end{equation}
and can be evaluated in terms of the standard elliptic integrals:
\begin{equation}\label{j-string-action-elliptic}
 S=-i\sqrt{\lambda\,\frac{1-j^2}{1-\kappa ^2}}
 \left(E(is_*)-F(is_*)\right)-\sqrt{\lambda\epsilon }\,.
\end{equation}
Instead of (\ref{sstar}), the argument of the elliptic functions is given by 
\begin{equation}\label{sstar-j}
 \cosh  s_*=\frac{1}{\kappa }\,\sqrt{\frac{1-j^2\kappa ^2}{1-j^2}}\,.
\end{equation}

A truly novel feature of the solution is an extra integration constant. We now need two boundary conditions to determine both
 $\epsilon $ and $j$ in terms of the physical parameters of the problem. The condition that the string reaches $z_0$ is similar to (\ref{z0}):
\begin{equation}\label{z0/R-new}
 \frac{z_0}{R}=\frac{\,{\rm e}\,^{v_*}}{\sqrt{1+\epsilon +j^2}}\,,
\end{equation}
where now
\begin{equation}\label{v-j-integral-form}
 v_*=\sqrt{\epsilon }\int_{\sqrt{\epsilon+j^2 }}^{\infty }\frac{dg}{(1+g^2)\sqrt{(1-j^2)g^2+g^4-\epsilon -j^2}}\,.
\end{equation}
This is derived by the same manipulations as (\ref{v-integral-form}) and integrates to
\begin{equation}\label{v-j-elliptic}
 v_*=i\sqrt{\frac{(1-j^2\kappa ^2)(\kappa ^2-j^2)}{(1-\kappa ^2)(1-j^2)}}\left(
 \Pi \left(\frac{1-\kappa ^2}{1-j^2}\,,is_*\right)-F(is_*)
 \right).
\end{equation}

The angle subtended by the string on $S^5$ should be equal to $\phi $. This is the second condition. Since the angle evolves linearly with $\sigma $: $\theta =j\sigma $, the total angle is given by
\begin{equation}\label{phi-int}
 \phi =j\int_{\sqrt{\epsilon +j^2}}^{\infty }\frac{dg}{\sqrt{(1-j^2)g^2+g^4-\epsilon -j^2}}\,.
\end{equation}
It can be also expressed as an elliptic integral:
\begin{equation}
 \phi =-ij\sqrt{\frac{1-\kappa ^2}{1-j^2}}\,F(is_*).
\end{equation}
This result, along with (\ref{z0/R-new}), (\ref{z0/R-new}), (\ref{v-j-elliptic}) and (\ref{j-string-action-elliptic}) expresses the string action as a function of the radius (or the ratio $z_0/R$) and the angle $\phi $ in a parametric form.

\begin{figure}[t]
 \centerline{\includegraphics[width=8cm]{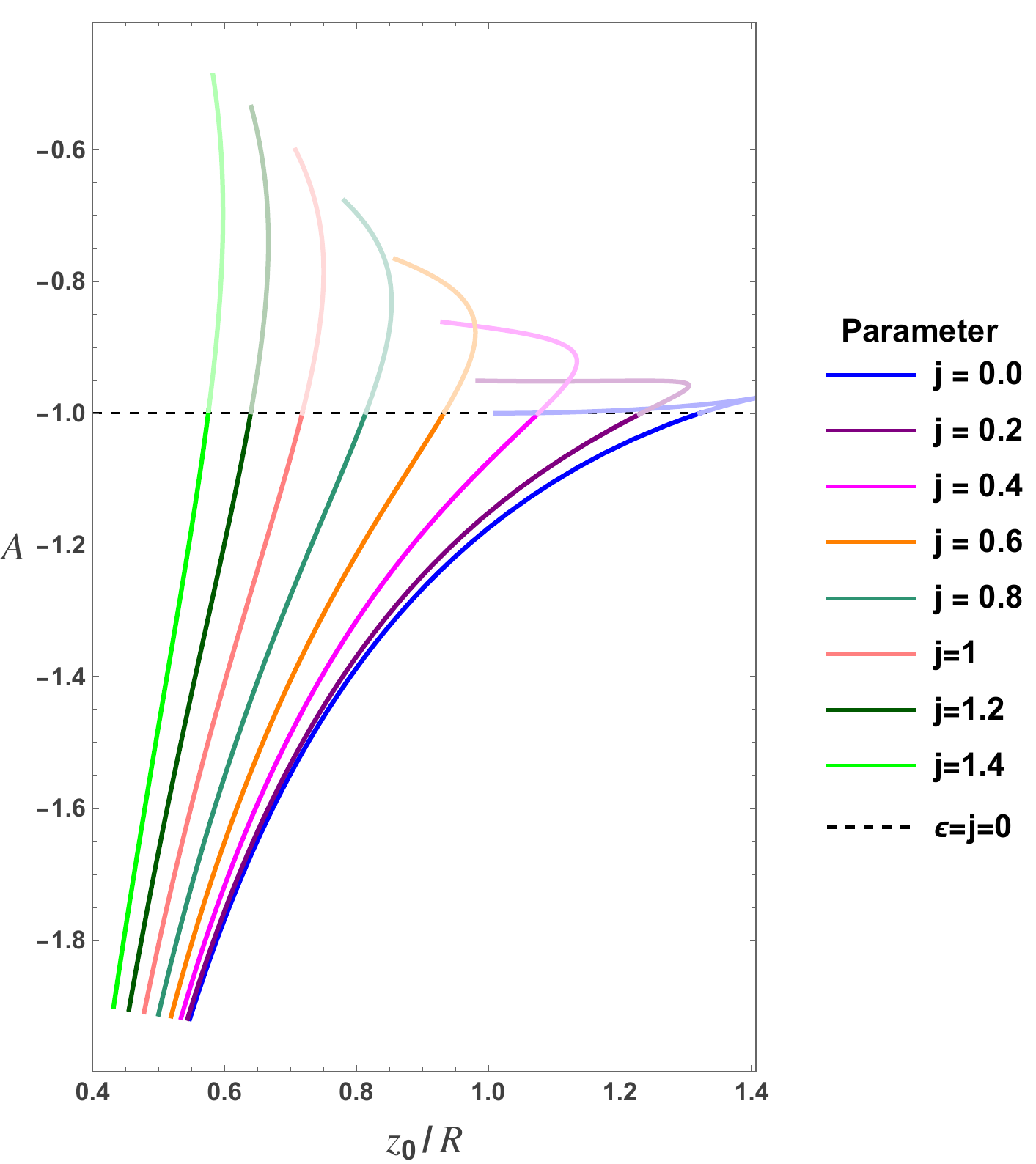}}
\caption{\label{Area}\small The area of the connected minimal surface for various values of $j$. The dashed line is the area of the disconnected surface:  in the shaded region above this line  the true minimum of the string action is the hemisphere  solution.}
\end{figure}

The solution has roughly the same structure as the minimal surface with $j=0$. In some range of parameters the solution has two branches. This range changes with the angle as illustrated in fig.~\ref{Area}. The area of the connected minimal surface always crosses $-1$,  undergoing the Gross-Ooguri transition. However, at non-zero $j$ the swallow-tail feature where the stable and unstable branches meet is rounded up. The transition occurs for smaller $z_0/R$ (larger radii of the circle) with growing $j$. This behavior is intuitively clear. The larger $j$ correspond to larger angles moving the contour  away from the brane in  the full $AdS_5\times S^5$ geometry. For keeping ten-dimensional distance constant the radial $AdS_5$ separation $z_0/R$ should be effectively smaller.

\begin{figure}[t]
 \centerline{\includegraphics[width=6cm]{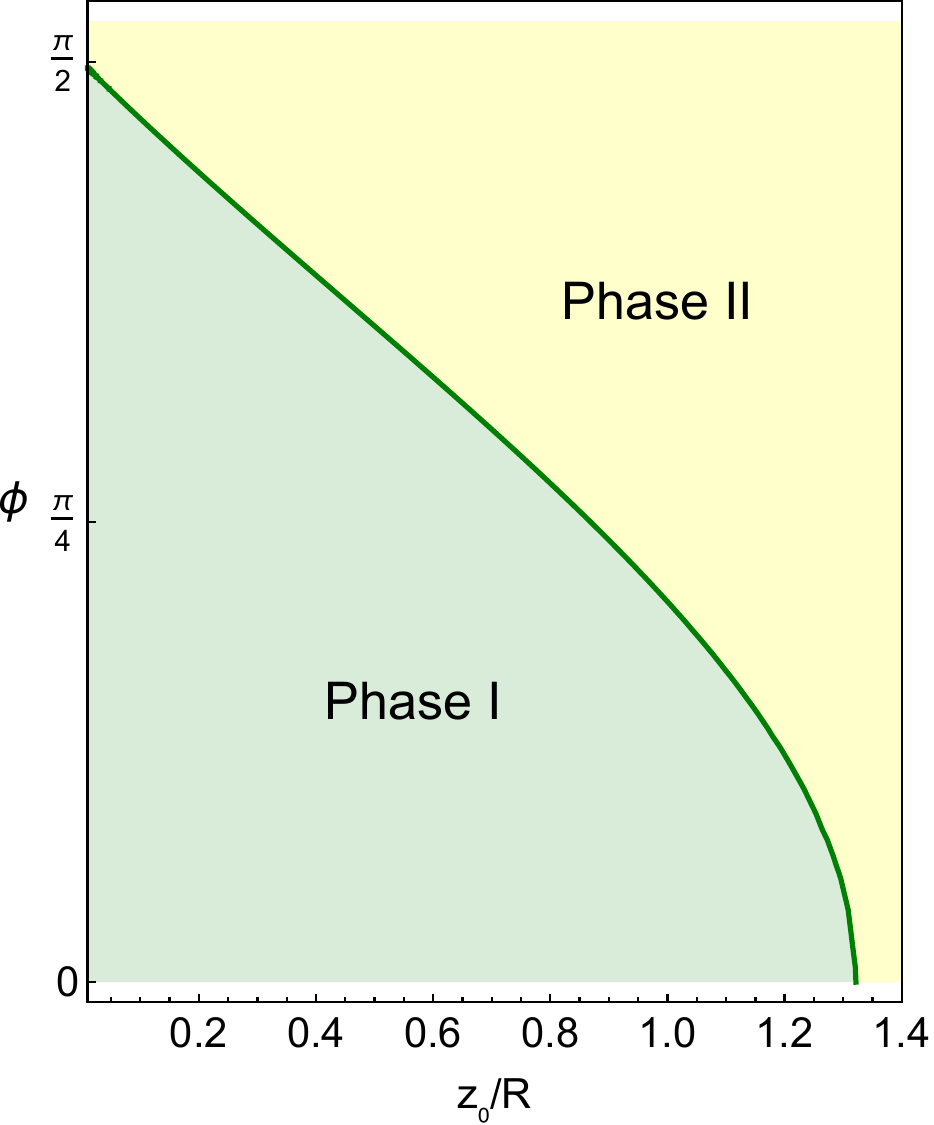}}
\caption{\label{p-diagram}\small The phase diagram of the Wilson loop in the angle-distance plane. The hemisphere is the dominant saddle-point in the phase II, while in the phase I the connected minimal surface has smaller area.}
\end{figure}

For large enough angular separation the connected surface does not even exist. On the region of integration in (\ref{phi-int}),
$$
(1-j^2)g^2+g^4-\epsilon -j^2=g^2(g^2-\epsilon -j^2)+(1+\epsilon )g^2-\epsilon -j^2>g^2(g^2-\epsilon -j^2).
$$
Hence,
\begin{equation}
 \phi < j\int_{\sqrt{\epsilon +j^2}}^{\infty }\frac{dg}{g\sqrt{g^2-\epsilon -j^2}}
 =\frac{\pi j}{2\sqrt{\epsilon +j^2}}<\frac{\pi }{2}\,.
\end{equation}
Hence, for $\phi >\pi /2$ the connected solution does not exists and the hemisphere is the only saddle-point of the string path integral. The full phase diagram is shown in fig.~\ref{p-diagram}.

The formulas simplify in the large-radius limit, which corresponds to the scaling regime of $\epsilon \sim j^2\gg 1$. The integrals in (\ref{phi-int}) and  (\ref{string-action-integral-j}) then evaluate to elementary functions, and $v_*$ in (\ref{z0/R-new}) can be neglected:
\begin{equation}
 z_0\simeq  \frac{R}{\sqrt{\epsilon +j^2}}\,,
 \qquad 
 \tan\phi \simeq \frac{j}{\sqrt{\epsilon }}\,,
 \qquad 
 S\simeq -\sqrt{\lambda \epsilon }\,, 
\end{equation}
or
\begin{equation}
 S=-\sqrt{\lambda }\,\,\frac{R\cos\phi }{z_0}=-2\pi Rv\cos\phi ,
\end{equation}
which reproduces the perimeter law (\ref{simple-straight}) for arbitrary angle $\phi <\pi /2$. The shape of the transition line near the corner is given by
\begin{equation}
 \frac{z_0}{R_{\rm c}}\simeq \cos\phi =\frac{\pi }{2}-\phi +\ldots 
\end{equation}
as  can be clearly seen in fig.~\ref{p-diagram}. The transition is always first order.

\section{Conclusions}

We studied classical string solutions that describe Wilson loops  of the simplest possible shape, the straight line and the circle. Albeit the solution for the circle is quite intricate, it can be found fully analytically. We believe that the underlying reason is integrability preserved by the D3-brane \cite{Demjaha:2025axy}.  We have not used  integrability in our analysis, and it would be interesting to make contact between the solution we found and integrable structures on the string worldsheet. 

The circular Wilson loop on the Coulomb branch features an intricate phase diagram that we mapped. Depending on the parameters, the minimal surface can extend all the way to the brane, or can  connect  to it through an infinitely thin tube. The two regimes are separated by a first order transition of Gross-Ooguri type. As in other examples of the Gross-Ooguri transition, this is a specifically holographic feature that should disappear at finite coupling \cite{Zarembo:1999bu}.

Quite remarkably we found that for a circle of a large radius the minimal area law results in a perturbative, BMN-like expansion derived directly  from string theory. It would be very interesting to recover this result by resummation of the planar diagrams, as has been done in similar contexts \cite{Aguilera-Damia:2016bqv,Bonansea:2020vqq,Kristjansen:2024map}. We also found evidence for non-renormalization of the straight Wilson line which we checked at the first non-trivial order at weak coupling but for which do not have a formal general proof.

\subsection*{Acknowledgements}

We would like  to thank C.~Kristjansen and C.~Uhlemann for interesting
discussions.
The work of K.~Z. was supported by VR grant 2021-04578. 

\appendix
\section{Shape of the minimal surface}\label{explicit-elliptic:sec}

\subsection{Aligned confirguration}

The differential equation (\ref{energy-conservation})  can be neatly integrated by a change of variables:
\begin{equation}\label{g-to-s}
 g=\frac{1}{\sqrt{1-\kappa ^2}\,\sinh s}\,,
\end{equation}
where $\kappa $ is defined in  (\ref{kappa-epsilon}), (\ref{epsilon-kappa}). For $\kappa =0$  the solution is (\ref{hemi-g}) and $s$ simply coincides with $\sigma $. More generally $s$ and $\sigma $ are related by 
\begin{equation}
 \frac{ds}{\sqrt{1-\kappa ^2\sinh^2s}}=\frac{d\sigma }{\sqrt{1-\kappa ^2}}\,.
\end{equation}
The left-hand side is an elliptic integral in its canonical form. The function $s(\sigma )$ is its inverse, the elliptic amplitude:
\begin{equation}\label{amplitude-ell}
 s=-i\mathop{\mathrm{am}}\frac{i\sigma }{\sqrt{1-\kappa ^2}}\,.
\end{equation}
The solution for $g$ in (\ref{g-to-s}) is consequently the elliptic cosecant:
\begin{equation}
 g=\frac{i}{\sqrt{1-\kappa ^2}}\,\mathop{\mathrm{ns}}\frac{i\sigma }{\sqrt{1-\kappa ^2}}\,.
\end{equation}

To find the shape of the minimal surface the dilaton equation  (\ref{dilaton}) needs to be integrated. This yields $\alpha $:
\begin{equation}\label{solution-for-alpha}
 \alpha =\frac{\kappa }{1-\kappa ^2}\int_{}^{}\frac{d\sigma }{1+g^2}-\ln\sqrt{1+g^2}\equiv v-\ln\sqrt{1+g^2}\,.
\end{equation}
The function denoted here by $v$ is expressed through elliptic integrals, and can be brought to the canonical form by the magic change of variables (\ref{g-to-s}):
\begin{equation}
 v=\frac{\kappa }{\sqrt{1-\kappa ^2}}\int_{}^{}ds\,
 \left[1-\frac{1}{1+(1-\kappa ^2)\sinh^2s}\right]\frac{1}{\sqrt{1-\kappa ^2\sinh^2s}}\,,
\end{equation}
and hence
\begin{equation}
 v=\frac{i\kappa }{\sqrt{1-\kappa ^2}}\left(
 \Pi (1-\kappa ^2,is)-F(is)
 \right).
\end{equation}
Then (\ref{g-alpha}) become:
\begin{align}
 &z=\frac{R\,{\rm e}\,^{v}\sinh s}{\sqrt{\cosh^2s+\frac{\kappa ^2}{1-\kappa ^2}}}\,,
\nonumber \\
&r=\frac{R\,{\rm e}\,^{v}}{\sqrt{(1-\kappa ^2)\cosh^2s+\kappa ^2}}\,.
\end{align}
These equations give an explicit parameterization of the minimal surface.
If we want to express the solution in the conformal coordinates
the hyperbolic functions of $s$ get replaced by the Jacobi functions of $\sigma $, in virtue of (\ref{amplitude-ell}):
\begin{equation}
 \sinh s=-i\mathop{\mathrm{sn}}\frac{i\sigma }{\sqrt{1-\kappa ^2}}\,,
 \qquad 
 \cosh s= \mathop{\mathrm{cn}}\frac{i\sigma }{\sqrt{1-\kappa ^2}}\,.
\end{equation}

\subsection{Misaligned configuration}

The solution with non-zero $j$ is still expressed through the elliptic functions. The magic change of variables, that brings them to the canonical form, is now
\begin{equation}
 g=\sqrt{\frac{1-j^2}{1-\kappa ^2}}\,\,\frac{1}{\sinh s}\,.
\end{equation}
With $s$ and $\sigma $ related by
\begin{equation}
 \frac{ds}{\sqrt{1-\kappa ^2\sinh^2s}}=\sqrt{\frac{1-j^2}{1-\kappa ^2}}\,\,d\sigma ,
\end{equation}
or equivalently
\begin{equation}
 s=-i\mathop{\mathrm{am}}\left(i\sigma \sqrt{\frac{1-j^2}{1-\kappa ^2}}\right),
\end{equation}
the solution for $g$ reads:
\begin{equation}
 g=i\sqrt{\frac{1-j^2}{1-\kappa ^2}}\,\mathop{\mathrm{ns}}
 \left(i\sigma \sqrt{\frac{1-j^2}{1-\kappa ^2}}\right).
\end{equation}

The solution of the dilaton equation (\ref{solution-for-alpha}) still holds, with the function $v$ given by  
\begin{equation}
 v=i\sqrt{\frac{(1-j^2\kappa ^2)(\kappa ^2-j^2)}{(1-\kappa ^2)(1-j^2)}}\left(
 \Pi \left(\frac{1-\kappa ^2}{1-j^2}\,,is\right)-F(is)
 \right),
\end{equation}
and the shape of the minimal surface is described by the two equations:
\begin{align}
 &z=\frac{R\,{\rm e}\,^{v}\sinh s}{\sqrt{\cosh^2s+\frac{\kappa ^2-j^2}{1-\kappa ^2}}}\,,
\nonumber \\
 &r=R\,{\rm e}\,^{v}\,\sqrt{\frac{1-j^2}{(1-\kappa ^2)\cosh^2s+\kappa ^2-j^2}}\,.
\end{align}
In the conformal coordinates,
\begin{equation}
 \sinh s=-i\mathop{\mathrm{sn}}\left(i\sigma \sqrt{\frac{1-j^2}{1-\kappa ^2}}\right),
 \qquad 
 \cosh s= \mathop{\mathrm{cn}}\left(i\sigma \sqrt{\frac{1-j^2}{1-\kappa ^2}}\right).
\end{equation}

Finally, the angle on $S^5$ is given by
\begin{equation}
 \theta  =-ij\sqrt{\frac{1-\kappa ^2}{1-j^2}}\,F(is).
\end{equation}

\bibliographystyle{nb}
\bibliography{refs}

\end{document}